\documentclass[aps,prl,twocolumn,superscriptaddress,showpacs]{revtex4}

\usepackage{graphicx}% Include figure files
\usepackage{dcolumn}% Align table columns on decimal point
\usepackage{bm}% bold math

\def\qsgw{\mbox{QS$GW$}}

\begin{document}

\title{Many-body Electronic Structure of Metallic $\bm{\alpha}$-Uranium}

\author{Athanasios N. Chantis} 
\affiliation{Theoretical Division, Los Alamos National Laboratory,
Los Alamos, New Mexico, 87545, USA}

\author{R. C. Albers} 
\affiliation{Theoretical Division, Los Alamos National Laboratory,
Los Alamos, New Mexico, 87545, USA}

\author{M. D. Jones} 
\affiliation{University at Buffalo, SUNY, Buffalo, New York 14260, USA}

\author{Mark van Schilfgaarde}
\affiliation{School of Materials, Arizona State University,
Tempe, Arizona, 85287-6006, USA}

\author{Takao Kotani}
\affiliation{School of Materials, Arizona State University,
Tempe, Arizona, 85287-6006, USA}

\date{\today}

\begin{abstract}

We present results for the electronic structure of $\alpha$ uranium
using a recently developed quasiparticle self-consistent $GW$ method
(\qsgw).  This is the first time that the $f$-orbital electron-electron
interactions in an actinide has been treated by a first-principles method
beyond the level of the generalized gradient
approximation (GGA) to the local density approximation (LDA).  
We show that the \qsgw\ approximation
predicts an $f$-level shift upwards
of about 0.5 eV with respect to the other metallic $s$-$d$ states
and that there is a significant $f$-band narrowing when
compared to LDA band-structure results. 
Nonetheless, because of the overall low $f$-electron occupation number
in uranium, ground-state properties and the occupied band structure
around the Fermi energy is not significantly affected.
The correlations predominate in the unoccupied part of the $f$ states.
This provides the first formal justification for the
success of LDA and GGA calculations in describing the ground-state
properties of this material.

%We show that a self-consistent form of the $GW$ method can successfully
%describe the electronic structure of materials that have both moderately
%correlated $spd$ and strongly correlated $f$ electrons.  Self-consistency
%is essential.  We predict a new effect, a first order metal-insulator
%transition in GdN, driven by dielectric function changes.  Because the
%method is rigorously grounded in many-body perturbation theory without any
%special ansatz, it can be used to examine approximations in some standard
%semiempirical approaches.

\end{abstract}

\pacs{71.15-m,71.10-w,71.20-Eh}

\maketitle

% ----------------- Start ---------------------

It has long been recognized that $5f$ electron-electron correlations
play an important role in the light actinides \cite{Lander03,actinide-book},
becoming increasingly significant as one moves across
this series and the atomic number Z increases.
This culminates in Pu, which has many extreme physical properties
that are driven by these correlations  \cite{Albers01},
such as the large volume expansion for the $\alpha$ to $\delta$ phase
transformation \cite{Hecker04a,Hecker04b}.
What is less clear is the role of correlations
for Z's less than Pu.
Uranium stands at a kind of threshold in this
regard.  Experimentally, the pure material is weak to moderately correlated \cite{Lander94},
since specific heat enhancements are moderate and no convincing satellite
or Kondo photoemission peaks are observed, which is consistent with the
success of band-structure in predicting materials properties \cite{Jones00,Soderlind02}.
At the same time, when the uranium atoms are pushed apart by other
elements, they form many heavy fermion and other strongly
correlated uranium compounds \cite{Fisk95}.
In this regard, uranium is an inviting target to study, since it should have
interesting correlation effects beyond conventional metals like copper
or aluminum, and yet these should be weak enough to have some hope
of accurately calculating them.  It is thus an important testing ground for correlation
theory and how many-body effects correct conventional LDA band structures.
    
The most widely used electronic-structure method, the local density
approximation (LDA), has been an immensely successful tool that
reasonably predicts ground-state properties of weakly correlated
systems.  The LDA is much less successful at predicting optical
properties of such systems, and its failures become more serious as
correlations become stronger.  Recent photoemission
spectroscopy on high quality uranium singe crystals has revealed
additional information about the electronic structure of this material
\cite{Opeil06,Opeil07}.  Comparison with LDA calculated electronic
bands shows some disagreement between experiment and theory.  Because
of the poor treatment of electron correlations by LDA it is
difficult to conclude whether the observed discrepancies between the
predicted band structure and photoemission data are due to electronic
correlations, even though perhaps weak,
or to other effects such as surface states. For the same
reason it is not clear how much of the mass enhancement 
observed in the specific heat \cite{Lashley01,Opeil06} can
be attributed to electron correlations and how much to electron-phonon
coupling.  To date all first-principles theoretical treatments of
the uranium electronic structure have been based on LDA or
the generalized-gradient approximation, GGA, extension to LDA. 
Therefore, it is important to explore the electronic structure of
uranium with methods that treat more accurately the electron-electron
interactions and to understand how they affect the electronic properties
of this material.

In Pu, it is now standard to use dynamical mean-field
theory (DMFT) to treat the strong correlations that go well beyond conventional
band-structure \cite{pu-dmft}.  However, this has the
unsatisfactory aspect that a model Hamiltonian is grafted onto
a band-structure approach in an ad hoc manner.
Because of the much weaker correlations in uranium,
it is possible in this material to instead use a more approximate
treatment of correlation effects that is completely first principles
and yet goes significantly beyond conventional band theory.
Thus, in this Letter we apply for first time for any actinide
a rigorous {\it ab initio} self-consistent many-body theoretical
approach, the $GW$ approximation, and show how electronic correlation
effects modify the electronic structure of uranium that is predicted by
LDA band theory.

In this work we use the recently developed $QSGW$ version \cite{kotani,MarkPRB06,Mark}
of the GW method, which itself can be viewed 
as the first term in the expansion 
of the non-local energy dependent self-energy $\Sigma(\bf{r},\bf{r},\omega)$
in the screened Coulomb interaction $W$. From a more physical point of view
the GW approximation can be interpreted as a dynamically screened Hartree-Fock approximation plus
a Coulomb hole contribution \cite{Hedin}.  Therefore, $GW$ is a 
well defined perturbation theory.  In its usual implemention, sometimes called
the "one-shot" approximation,
it depends on an one-electron Green's functions based on LDA eigenvalues and eigenfunctions,
and the results can depend on this choice.
We have demonstrated~\cite{MarkPRB06} that
as correlations become stronger
serious practical and formal problems can arise in this approximation.
In Ref.~\cite{kotani} a formal description
is provided on how \qsgw\ is a rigorous
way to surmount this difficulty, based on using a self-consistent
one-electron Green's function that is derived from the self-energy
(the quasi-particle eigenvalues and eigenfunctions) instead of an LDA starting point.
In the literature, it is has been demonstrated that \qsgw\
version of GW theory reliably 
describes a wide range of
$spd$ systems~\cite{Mark,Faleev,chantis} and rare-earths~\cite{fshells}.
It should be noted that the energy eigenvalues of the QSGW method are the same as the quasiparticle
spectra of the GW method (i.e., the peaks in the self-energy).  This captures
the many-body shifts in the quasiparticle energies.  However, when presenting the
quasiparticle DOS this ignores the smearing by the imaginary part of the self-energy
of the spectra due to quasiparticle lifetime effects, which should increase as one
moves farther away from the Fermi energy.

The \qsgw\ method is currently implemented using a generalization of the
Full Potential Linear Muffin Tin Orbital (FP-LMTO) method \cite{markbook},
so we make no approximations for the shape of crystal potential.  The
smoothed LMTO basis\cite{MarkPRB06} includes orbitals with $l \le
l_{max}=6$; both 7$p$ and 6$p$ as well as both 5$f$ and 6$f$ are included
in the basis. $6f$ are added in the form of local orbitals
\cite{MarkPRB06}, that is an orbital strictly confined to the augmentation
sphere, and has no envelope function at all.  $7p$ are added as a kind of
extended local orbitals the 'head' of which is evaluated at an energy far
above Fermi level \cite{MarkPRB06}, instead of making the orbital vanish at
the augmentation radius a smooth Hankel `tail' is attached to the orbital.
A particularly important point is that core states are treated at the
exchange-only level.  We have demonstrated in some
detail~\cite{Kotani02,MarkPRB06} that approximating the core by the LDA
potential, i.e. computing $\Sigma$ from the valence electrons only, sometimes
leads to significant errors. Since \qsgw\ gives the self-consistent solution at the scalar relativistic
level, we add the spin-orbit operator
$H_{SO}=\mathbf{L}\cdot\mathbf{S}/2c^{2}$ as a perturbation (it is not
included in the self-consistency cycle).  For our calculations we use the
equilibrium crystal structure of $\alpha$-U, the orthorhombic Cmcm, with the
Uranium atoms located at the 4c positions:(0,$y$,$\frac{1}{4}$) and
(0,-$y$,$\frac{3}{4}$) plus C centering; we use the experimental lattice
parameters $a=2.858$\AA, $b=5.876$\AA, $c=4.955$\AA, and $y$=0.105.

\begin{figure}[tbp]
\includegraphics[angle=0,width=0.45\textwidth,clip]{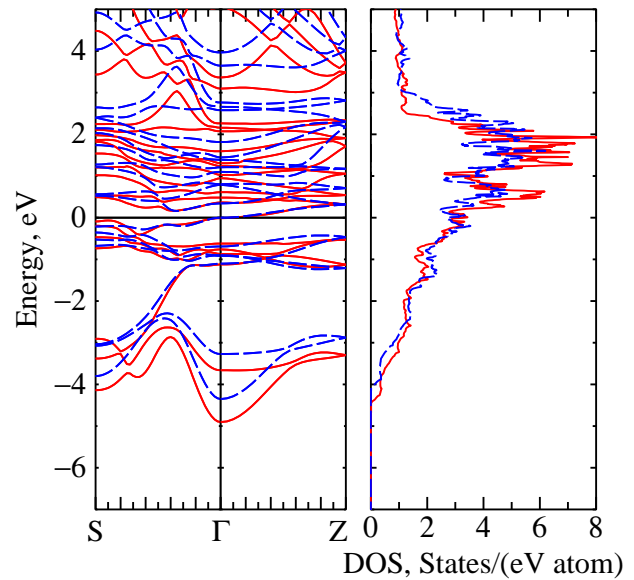}
\caption{ The energy bands (or quasi-particle energies) along two
symmetry directions (left panel) and the total DOS (right panel);
the \qsgw\ results are represented by solid red lines and the LDA
band-structure results by dashed blue lines.}
\label{fig:band}
\end{figure}
%%%%%%%%%%%%%%%%%%%%%%%%%%%%%%%%%%%%%%%%%%%%%%%%%%%%%%%%%%

\begin{figure}[tbp]
\includegraphics[angle=0,width=0.45\textwidth,clip]{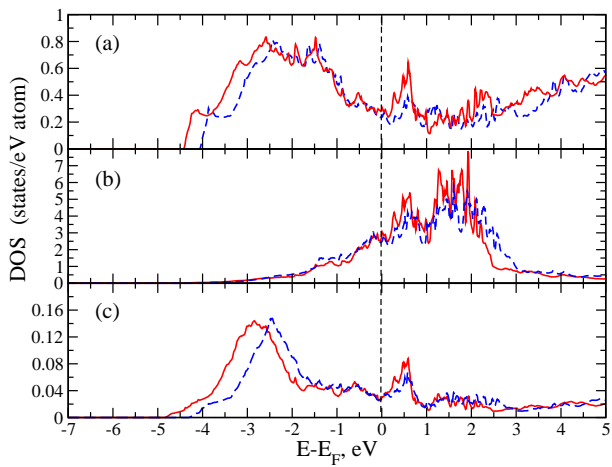}
\caption{ \small (a) The $d$ partial DOS (b) The $f$ partial DOS (c) The $s$ partial DOS. 
In all cases the \qsgw\ results are the solid red lines, and the LDA results
are the dashed blue lines.}
\label{fig:pdos}
\end{figure}

In Fig.~1 we compare the calculated \qsgw\ one-particle electronic
structure of $\alpha$ U with the LDA band-structure results.  In both
cases, the narrow bands located approximately between -1 and +3 eV are mainly due to
the uranium 5$f$ orbitals; the lowest dispersive bands seen on this plot
have $s$ character; and the bands above 3 eV are strongly
hybridized.  Both methods are roughly in agreement for the total
density of states (DOS).  However, the large DOS peak that is a little
above the Fermi energy, $E_{F}$, is narrower in width and larger in
magnitude for the \qsgw\ calculation; also,
the quasiparticle energies only agree well with the LDA band-structure
results in the vicinity of
$E_{F}$.  As we move to higher or lower energies (away from $E_F$) the
difference between \qsgw\ and LDA quasiparticle energies gradually
increase.  Among the occupied states, the metallic $s$-$d$ bands at the lowest
energies experience a significant downward shift relative to the $f$
bands when compared to the LDA results
(note that the main part of the $p$ states are believed to lie above $E_F$,
since they are repelled by the 6$p$ semi-core states that fall below
and are well separated from
the conduction band, and hence only $p$ hybridization tails
appear in the occupied conduction-band region).  For example, at the
$\Gamma$ point the shift is about 1 eV downwards, but more generally,
however,  the energy shift is somewhat $\mathbf{k}$-dependent.  
The partial DOS presented in Fig.~\ref{fig:pdos} shows that after
integration over all $\mathbf{k}$ there is a downward energy shift
of about 0.5 eV for the occupied $s$-$d$ bands.

%\begin{figure}[tbp]
%\includegraphics[angle=0,width=0.45\textwidth,clip]{alpha-U-fig3.eps}
%\includegraphics[angle=0,width=0.25\textwidth]{/home/chantis/lm-fp/d-gw-lda.ps}
%\includegraphics[angle=0,width=0.25\textwidth]{/home/chantis/lm-fp/s-gw-lda.ps}
%\includegraphics[angle=0,width=0.23\textwidth,clip]{eras-bnds-dos.eps}
%\caption{ \small (a) Comparison LAPW LDA and LAPW LDA calculations (b)Comparison of LMTO and LAPW LDA calculations. }
%\label{fig:lapw}
%\end{figure}
In Fig.~\ref{fig:pdos}(b) we present
the partial DOS for the $f$ bands.  One of the important effects of
electronic correlation is to narrow the width of a band.  This shows
up as a narrower $f$-band width in the $QSGW$ calculation. In
addition, since the area under the curve is proportional to the number
of $f$ states, which remains constant, the amplitude of the
quasiparticle peak is also higher.  The narrowing of the $f$ band
together with its energy shift results in a slight change of the
electron occupation. Comparison of the partial DOS shows that  
the $f$ band shifts up, relatively to $s$ and $d$ bands, toward unoccupied states. 
The $f$ occupation is equal to 3.19 in the $QSGW$
and 3.57 in the LDA calculation, hence in $QSGW$ about 0.4 $f$ electron is 
lost to the $s$-$d$ interstitial charge. The overall $f$ occupation in uranium
is relatively low so that the position of the Fermi level remains in
the lower part of the $f$ peak, where the difference between $QSGW$
and LDA calculation is negligible.  For this reason, even though the
$5f$ electron states in uranium appear to be correlated, the physical
properties that are related only to occupied electron states should be
predicted well by the LDA approximation.  It is mainly in the excited-state
spectra of the $f$ states above the Fermi energy where the correlation
effects are strongly apparent. Consistent with the weak to moderate strength of the
correlations, we find that the first iteration of the $QSGW$ method, which is sometimes
called the "one-shot GW" is very similar to the fully self-consistent $QSGW$ results.

The electron DOS at the Fermi level in $QSGW$ is 3.35 states/(eV atom) while in
the LDA result is 2.75 states/(eV atom).  We have found that the DOS has a lot of
variation around the Fermi level and requires a very large number
of k points to converge (we used $82^{3}$).  This explains deviations of our
DOS with earlier results (e.g., Ref.~\onlinecite{Opeil07} and references therein)
Our LDA LMTO results using a Barth-Hedin
exchange-correlation potential were found to be in excellent agreement
with GGA (gradient-corrected) LAPW  calculations using a PBE exchange-correlation
potential that we calculated with the Wien2k code \cite{wien2k-01};
e.g., we found insignificant differences in the total LDA and GGA DOS.
 %In our self-consistent $GW$
%method we construct at each iteration step $n$ an effective exchange
%correlation potential $V^{n}_{xc}=(1/2) \sum | \psi_i \rangle
%Re[(\Sigma(\epsilon_{i})]_{ij}+Re[\Sigma(\epsilon_{j})]_{ij} \langle
%\psi_{j}|$, where the next iteration of the Hamiltonian is generated from
%$H^{n}=H^{n-1}+V^{n}_{xc}-V^{n-1}_{xc}$. At self-consistency
%$V^{n}_{xc}=V^{n-1}_{xc}$. 

It was shown by Luttinger \cite{Luttinger} that
the linear coefficient of specific heat for a system of interacting
electrons is given by the quasiparticle DOS, i.e., $\gamma \sim
\sum_{\mathbf{k}} \delta(E_F-E(\mathbf{k}))$.  This can be compared to
the one-electron coefficient, which is given by $\gamma^0 \sim
\sum_{\mathbf{k}} \delta(E_F^0-\epsilon(\mathbf{k}))$.  Here, $E_F$
and $E(\mathbf{k})$ are the Fermi and quasiparticle energies of the
interacting electron gas (in the $QSGW$ approximation), 
while $E_F^0$ and $\epsilon(\mathbf{k})$ are
the Fermi energy and band-structure eigenvalues.  Hence
only the quasiparticle shifts are needed for calculating the
specific heat, which are included by construction in the
QSGW energy eigenvalues and DOS. Therefore, the specific
heat of the interacting \qsgw\ electron gas is proportional to the
\qsgw\ DOS at the Fermi level $N(E_F)$, and there is no need to
include an additional renormalization factor $(1-d\Sigma/d\omega)$ factor, which
in model calculations converts the band-structure DOS to the
quasiparticle DOS.  We find that
%$N^{QSGW}(E_{F})/N^{LDA}(E_{F})=\gamma^{QSGW}/\gamma^{LDA} = 1.22$. From the DOS at
the linear coefficient of specific heat
in \qsgw\ is $\gamma$= 7.89 mJ mol$^{-1}$ K$^{-2}$ while in LDA is $\gamma$= 6.48 mJ mol$^{-1}$ K$^{-2}$,
giving a \qsgw\ enhancement factor of $N^{QSGW}(E_{F})/N^{LDA}(E_{F})=\gamma^{QSGW}/\gamma^{LDA} = 1.22$.
%$\gamma_{QSGW}$=0.793$\times$10$^{-3}${}$R$K$^{-1}$.
A recently measured value \cite{Lashley01},
%$\gamma_{exp}$=1.10$\times$10$^{-3}${}$R$K$^{-1}$
$\gamma_{exp}$=9.15{} mJ mol$^{-1}$ K$^{-2}$, is larger than the \qsgw\ result by a factor
of $\gamma_{exp}/\gamma_{QSGW}$=1.16 and larger than the LDA result by a factor of
$\gamma_{exp}/\gamma_{LDA}$=1.41.  The relatively small remaining enhancement that
is not accounted for by \qsgw\ must almost certainly be the electron-phonon enhancements
that are present in all metals and that are typically at least this big.

%It is worth mentioning that unlike the case of 
%open shell rare-earths studied in Ref.~\cite{fshells}, uranium's self-consistent $QSGW$
%band structure is very similar to that of the first 
%$QSGW$ iteration.
%This implies that $QSGW$ and the "one-shot" $GW$ methods give very similar results
%for uranium, probably due to the strong metallic screening in this system.
%Therefore, at least in the case of uranium, errors beyond \qsgw\ approximation are 
%the same with those beyond other versions of $GW$ methods, e.g.,
%ladder-diagram electron-hole interactions (similar to excitonic
%effects) and vertex corrections to $GW$ \cite{Hedin}. At present it is
%beyond our ability to include such effects explicitly in the
%calculations, but it is known that such effects mainly play an
%important role for systems that are much more strongly correlated 
%(e.g., creating sattelite peaks in photoelectron spectra).  
%Therefore, since our results suggest that $\alpha$ U is only
%moderately correlated, we believe that the observed
%experimental specific heat enhancement is probably due to electron-phonon
%interactions.

\begin{figure}[tbp]
\includegraphics[angle=0,width=0.45\textwidth,clip]{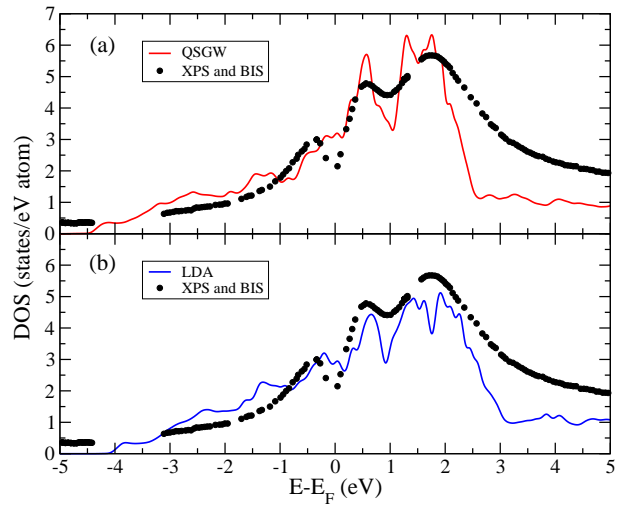}
\caption{ \small (a) Comparison of photemission data with the calculated \qsgw\ DOS 
(b)Comparison of photemission data with the calculated LDA DOS. In both cases, the DOS has a Gaussian broadening of 0.05 eV}
\label{fig:bis}
\end{figure}

The bremsstrahlung isochromat spectroscopy (BIS) of $\alpha$ uranium
exhibits a double peak structure within the interval of 1 to 2 eV
above $E_{F}$ \cite{Baer}; as one can see in Fig.~\ref{fig:bis}, this
feature is reproduced by both calculations. 
However, further comparison of the two methods with photoemission data 
is difficult because the photoemission peaks are too broad comparing to the
band width narrowing and quasiparticle shifts observed with the help of
$QSGW$ method. 
%placement of the second peak just below 2 eV is in better agreement
%with the BIS data.  This is an experimental confirmation of the
%quasiparticle band narrowing caused by the electronic correlations.  
Recent angle resolved photoemission (ARPES)\cite{Opeil07} 
and ultraviolet photoemission spectroscopy (UPS)\cite{Opeil06} were found to be 
in good agreement with GGA band structure calculations. 
However, a low energy UPS peak and several APRES local maxima
were not predicted by GGA band structure.  
Since the \qsgw\ energy bands along $\Gamma \to Z$ direction in the energy
window -2 to 0 eV are very similar to those obtained with LDA the
agreement with the valence band (UPS) is of the same level with LDA. The 
unexplained APRES local maxima are located in the vicinity of $\Gamma$ point 
with energies around -5 eV and -2 eV \cite{Opeil06}. Even though the
$QSGW$ bands along $\Gamma \to S$ direction agree less with LDA than they do
along $\Gamma \to Z$ direction, the differences are not significant enough
to explain the aforementioned ARPES data.
%So based on our $QSGW$ results, we note that
%the observed UPS peak just below $E_F$ \cite{Opeil06} and the
%is unlikely
%to be due to a quasiparticle shift of $5f$ bands; 
We suspect that
the presence of uranium surface states as was suggested in Ref.~\cite{Opeil07}
or final state effects are a more likely explanation for these features.

\begin{figure}[tbp]
\includegraphics[angle=0,width=0.3\textwidth]{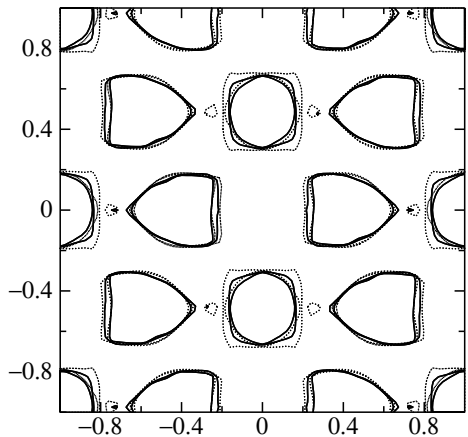}
\caption{ \small The Fermi surface cross section cut by the [100]-[010]
plane at h=0.288; \qsgw\ -- solid lines, LDA -- dotted lines.}
\label{fig:band}
\end{figure}

The [100]-[010] cross section of the Fermi surface is presented in
Fig.~3.  It shows that there are only slight changes in \qsgw\
compared to the LDA calculations. This is representative of several
cross sections that were plotted, showing that
overall the electron correlations have no significant effect on the
shape and size of the Fermi surface.

In conclusion, we have used the $QSGW$ method to show that moderate
$f$-electron correlation effects are present in $\alpha$-U, and that it
is because of the low occupation of $f$ electrons that these effects
don't show up more strongly in Fermi surface and other ground-state
properties of this material.  Most of the correlation effects only appear
in the excited-state spectra in the unoccupied $f$ states. As is commonly
suspected, LDA or GGA band-structure methods somewhat misplace
narrow bands, such as $d$ and $f$ bands, with respect to the remaining
metallic bands.  For uranium the error is about 0.5 eV.

This work was carried out under the auspices of the National 
Nuclear Security Administration of the U.S. Department of Energy 
at Los Alamos National Laboratory under Contract No. DE-AC52-06NA25396.
MvS and TK acknowledge support from ONR (contract N00014-07-1-0479) and 
by DOE (contract DE-FG02-06ER46302) and would like to thank the Fulton HPC 
for computational resources used in this project. 
ANC would like to thank Axel Svane for informative discussions.

\begin{acknowledgments}

\end{acknowledgments}

\bibliography{u-gw-rev}

\begin{thebibliography}{25}
\expandafter\ifx\csname natexlab\endcsname\relax\def\natexlab#1{#1}\fi
\expandafter\ifx\csname bibnamefont\endcsname\relax
  \def\bibnamefont#1{#1}\fi
\expandafter\ifx\csname bibfnamefont\endcsname\relax
  \def\bibfnamefont#1{#1}\fi
\expandafter\ifx\csname citenamefont\endcsname\relax
  \def\citenamefont#1{#1}\fi
\expandafter\ifx\csname url\endcsname\relax
  \def\url#1{\texttt{#1}}\fi
\expandafter\ifx\csname urlprefix\endcsname\relax\def\urlprefix{URL }\fi
\providecommand{\bibinfo}[2]{#2}
\providecommand{\eprint}[2][]{\url{#2}}

\bibitem[{\citenamefont{Lander}(2003)}]{Lander03}
\bibinfo{author}{\bibfnamefont{G.~H.} \bibnamefont{Lander}},
  \bibinfo{journal}{Science} \textbf{\bibinfo{volume}{301}},
  \bibinfo{pages}{1057} (\bibinfo{year}{2003}).

\bibitem[{\citenamefont{Freeman and Lander}(1984)}]{actinide-book}
\bibinfo{editor}{\bibfnamefont{A.~J.} \bibnamefont{Freeman}} \bibnamefont{and}
  \bibinfo{editor}{\bibfnamefont{G.~H.} \bibnamefont{Lander}}, eds.,
  \emph{\bibinfo{title}{Handbook on the Physics and Chemistry of the
  Actinides}}, vol. \bibinfo{volume}{1 and 2}
  (\bibinfo{publisher}{North-Holland, Amsterdam}, \bibinfo{year}{1984}).

\bibitem[{\citenamefont{Albers}(2001)}]{Albers01}
\bibinfo{author}{\bibfnamefont{R.~C.} \bibnamefont{Albers}},
  \bibinfo{journal}{Nature} \textbf{\bibinfo{volume}{410}},
  \bibinfo{pages}{759} (\bibinfo{year}{2001}).

\bibitem[{\citenamefont{Hecker}(2004)}]{Hecker04a}
\bibinfo{author}{\bibfnamefont{S.~S.} \bibnamefont{Hecker}},
  \bibinfo{journal}{Met. Mat. Trans. A} \textbf{\bibinfo{volume}{35A}},
  \bibinfo{pages}{2207} (\bibinfo{year}{2004}).

\bibitem[{\citenamefont{Hecker et~al.}(2004)\citenamefont{Hecker, Harbur, and
  Zocco}}]{Hecker04b}
\bibinfo{author}{\bibfnamefont{S.~S.} \bibnamefont{Hecker}},
  \bibinfo{author}{\bibfnamefont{D.~R.} \bibnamefont{Harbur}},
  \bibnamefont{and} \bibinfo{author}{\bibfnamefont{T.~G.} \bibnamefont{Zocco}},
  \bibinfo{journal}{Prog. Mat. Sci.} \textbf{\bibinfo{volume}{49}},
  \bibinfo{pages}{429} (\bibinfo{year}{2004}).

\bibitem[{\citenamefont{Lander et~al.}(1994)\citenamefont{Lander, Fisher, and
  Bader}}]{Lander94}
\bibinfo{author}{\bibfnamefont{G.~H.} \bibnamefont{Lander}},
  \bibinfo{author}{\bibfnamefont{E.~S.} \bibnamefont{Fisher}},
  \bibnamefont{and} \bibinfo{author}{\bibfnamefont{S.~D.} \bibnamefont{Bader}},
  \bibinfo{journal}{Adv. Phys.} \textbf{\bibinfo{volume}{43}},
  \bibinfo{pages}{1} (\bibinfo{year}{1994}).

\bibitem[{\citenamefont{Jones et~al.}(2000)\citenamefont{Jones, Boettger,
  Albers, and Singh}}]{Jones00}
\bibinfo{author}{\bibfnamefont{M.~D.} \bibnamefont{Jones}},
  \bibinfo{author}{\bibfnamefont{J.~C.} \bibnamefont{Boettger}},
  \bibinfo{author}{\bibfnamefont{R.~C.} \bibnamefont{Albers}},
  \bibnamefont{and} \bibinfo{author}{\bibfnamefont{D.~J.} \bibnamefont{Singh}},
  \bibinfo{journal}{Phys. Rev. B} \textbf{\bibinfo{volume}{61}},
  \bibinfo{pages}{4644} (\bibinfo{year}{2000}).

\bibitem[{\citenamefont{Soderlind}(2002)}]{Soderlind02}
\bibinfo{author}{\bibfnamefont{P.}~\bibnamefont{Soderlind}},
  \bibinfo{journal}{Phys. Rev. B} \textbf{\bibinfo{volume}{66}},
  \bibinfo{pages}{085113} (\bibinfo{year}{2002}).

\bibitem[{\citenamefont{Fisk et~al.}(1995)\citenamefont{Fisk, Sarrao, Smith,
  and Thompson}}]{Fisk95}
\bibinfo{author}{\bibfnamefont{Z.}~\bibnamefont{Fisk}},
  \bibinfo{author}{\bibfnamefont{J.~L.} \bibnamefont{Sarrao}},
  \bibinfo{author}{\bibfnamefont{J.~L.} \bibnamefont{Smith}}, \bibnamefont{and}
  \bibinfo{author}{\bibfnamefont{J.~D.} \bibnamefont{Thompson}},
  \bibinfo{journal}{Proc. Natl. Acad. Sci. USA} \textbf{\bibinfo{volume}{92}},
  \bibinfo{pages}{6663} (\bibinfo{year}{1995}).

\bibitem[{\citenamefont{Opeil et~al.}(2006)\citenamefont{Opeil, Schulze,
  Manley, Lashley, Hults, R.~J.~Hanrahan, Smith, Mihaila, Blagoev, Albers
  et~al.}}]{Opeil06}
\bibinfo{author}{\bibfnamefont{C.~P.} \bibnamefont{Opeil}},
  \bibinfo{author}{\bibfnamefont{R.~K.} \bibnamefont{Schulze}},
  \bibinfo{author}{\bibfnamefont{M.~E.} \bibnamefont{Manley}},
  \bibinfo{author}{\bibfnamefont{J.~C.} \bibnamefont{Lashley}},
  \bibinfo{author}{\bibfnamefont{W.~L.} \bibnamefont{Hults}},
  \bibinfo{author}{\bibfnamefont{J.}~\bibnamefont{R.~J.~Hanrahan}},
  \bibinfo{author}{\bibfnamefont{J.~L.} \bibnamefont{Smith}},
  \bibinfo{author}{\bibfnamefont{B.}~\bibnamefont{Mihaila}},
  \bibinfo{author}{\bibfnamefont{K.~B.} \bibnamefont{Blagoev}},
  \bibinfo{author}{\bibfnamefont{R.~C.} \bibnamefont{Albers}},
  \bibnamefont{et~al.}, \bibinfo{journal}{Phys. Rev. B}
  \textbf{\bibinfo{volume}{73}}, \bibinfo{pages}{165109}
  (\bibinfo{year}{2006}).

\bibitem[{\citenamefont{Opeil et~al.}(2007)\citenamefont{Opeil, Schulze, Volz,
  Lashley, Manley, Hults, R.~J.~Hanrahan, Smith, Mihaila, Blagoev
  et~al.}}]{Opeil07}
\bibinfo{author}{\bibfnamefont{C.~P.} \bibnamefont{Opeil}},
  \bibinfo{author}{\bibfnamefont{R.~K.} \bibnamefont{Schulze}},
  \bibinfo{author}{\bibfnamefont{H.~M.} \bibnamefont{Volz}},
  \bibinfo{author}{\bibfnamefont{J.~C.} \bibnamefont{Lashley}},
  \bibinfo{author}{\bibfnamefont{M.~E.} \bibnamefont{Manley}},
  \bibinfo{author}{\bibfnamefont{W.~L.} \bibnamefont{Hults}},
  \bibinfo{author}{\bibfnamefont{J.}~\bibnamefont{R.~J.~Hanrahan}},
  \bibinfo{author}{\bibfnamefont{J.~L.} \bibnamefont{Smith}},
  \bibinfo{author}{\bibfnamefont{B.}~\bibnamefont{Mihaila}},
  \bibinfo{author}{\bibfnamefont{K.~B.} \bibnamefont{Blagoev}},
  \bibnamefont{et~al.}, \bibinfo{journal}{Physical Review B}
  \textbf{\bibinfo{volume}{75}}, \bibinfo{eid}{045120} (\bibinfo{year}{2007}).

\bibitem[{\citenamefont{Lashley et~al.}(2001)\citenamefont{Lashley, Lang,
  Boerio-Goates, Woodfield, Schmiedeshoff, Gay, McPheeters, Thoma, Hults,
  Cooley et~al.}}]{Lashley01}
\bibinfo{author}{\bibfnamefont{J.~C.} \bibnamefont{Lashley}},
  \bibinfo{author}{\bibfnamefont{B.~E.} \bibnamefont{Lang}},
  \bibinfo{author}{\bibfnamefont{J.}~\bibnamefont{Boerio-Goates}},
  \bibinfo{author}{\bibfnamefont{B.~F.} \bibnamefont{Woodfield}},
  \bibinfo{author}{\bibfnamefont{G.~M.} \bibnamefont{Schmiedeshoff}},
  \bibinfo{author}{\bibfnamefont{E.~C.} \bibnamefont{Gay}},
  \bibinfo{author}{\bibfnamefont{C.~C.} \bibnamefont{McPheeters}},
  \bibinfo{author}{\bibfnamefont{D.~J.} \bibnamefont{Thoma}},
  \bibinfo{author}{\bibfnamefont{W.~L.} \bibnamefont{Hults}},
  \bibinfo{author}{\bibfnamefont{J.~C.} \bibnamefont{Cooley}},
  \bibnamefont{et~al.}, \bibinfo{journal}{Phys. Rev. B}
  \textbf{\bibinfo{volume}{63}}, \bibinfo{pages}{224510}
  (\bibinfo{year}{2001}).

\bibitem[{pu-()}]{pu-dmft}
\bibinfo{note}{See, for example, Jian-Xin Zhu, et al., Phys. Rev. B
  \textbf{76}, 245118 (2007) and references therein.}

\bibitem[{\citenamefont{Kotani et~al.}(2007)\citenamefont{Kotani, van
  Schilfgaarde, and Faleev}}]{kotani}
\bibinfo{author}{\bibfnamefont{T.}~\bibnamefont{Kotani}},
  \bibinfo{author}{\bibfnamefont{M.}~\bibnamefont{van Schilfgaarde}},
  \bibnamefont{and} \bibinfo{author}{\bibfnamefont{S.~V.}
  \bibnamefont{Faleev}}, \bibinfo{journal}{Phys. Rev. B}
  \textbf{\bibinfo{volume}{76}}, \bibinfo{pages}{165106}
  (\bibinfo{year}{2007}).

\bibitem[{\citenamefont{van Schilfgaarde
  et~al.}(2006{\natexlab{a}})\citenamefont{van Schilfgaarde, Kotani, and
  Faleev}}]{MarkPRB06}
\bibinfo{author}{\bibfnamefont{M.}~\bibnamefont{van Schilfgaarde}},
  \bibinfo{author}{\bibfnamefont{T.}~\bibnamefont{Kotani}}, \bibnamefont{and}
  \bibinfo{author}{\bibfnamefont{S.~V.} \bibnamefont{Faleev}},
  \bibinfo{journal}{Phys. Rev. B} \textbf{\bibinfo{volume}{74}},
  \bibinfo{pages}{245125} (\bibinfo{year}{2006}{\natexlab{a}}).

\bibitem[{\citenamefont{van Schilfgaarde
  et~al.}(2006{\natexlab{b}})\citenamefont{van Schilfgaarde, Kotani, and
  Faleev}}]{Mark}
\bibinfo{author}{\bibfnamefont{M.}~\bibnamefont{van Schilfgaarde}},
  \bibinfo{author}{\bibfnamefont{T.}~\bibnamefont{Kotani}}, \bibnamefont{and}
  \bibinfo{author}{\bibfnamefont{S.}~\bibnamefont{Faleev}},
  \bibinfo{journal}{Phys. Rev. Lett.} \textbf{\bibinfo{volume}{96}},
  \bibinfo{pages}{226402} (\bibinfo{year}{2006}{\natexlab{b}}).

\bibitem[{\citenamefont{Hedin}(1999)}]{Hedin}
\bibinfo{author}{\bibfnamefont{L.}~\bibnamefont{Hedin}}, \bibinfo{journal}{J.
  Phys.: Condens. Matter} \textbf{\bibinfo{volume}{11}}, \bibinfo{pages}{R489}
  (\bibinfo{year}{1999}).

\bibitem[{\citenamefont{Faleev et~al.}(2004)\citenamefont{Faleev, van
  Schilfgaarde, and Kotani}}]{Faleev}
\bibinfo{author}{\bibfnamefont{S.~V.} \bibnamefont{Faleev}},
  \bibinfo{author}{\bibfnamefont{M.}~\bibnamefont{van Schilfgaarde}},
  \bibnamefont{and} \bibinfo{author}{\bibfnamefont{T.}~\bibnamefont{Kotani}},
  \bibinfo{journal}{Phys. Rev. Lett.} \textbf{\bibinfo{volume}{93}},
  \bibinfo{pages}{126406} (\bibinfo{year}{2004}).

\bibitem[{\citenamefont{Chantis et~al.}(2006)\citenamefont{Chantis, van
  Schilfgaarde, and Kotani}}]{chantis}
\bibinfo{author}{\bibfnamefont{A.~N.} \bibnamefont{Chantis}},
  \bibinfo{author}{\bibfnamefont{M.}~\bibnamefont{van Schilfgaarde}},
  \bibnamefont{and} \bibinfo{author}{\bibfnamefont{T.}~\bibnamefont{Kotani}},
  \bibinfo{journal}{Phys. Rev. Lett.} \textbf{\bibinfo{volume}{96}},
  \bibinfo{pages}{086405} (\bibinfo{year}{2006}).

\bibitem[{\citenamefont{Chantis et~al.}(2007)\citenamefont{Chantis, van
  Schilfgaarde, and Kotani}}]{fshells}
\bibinfo{author}{\bibfnamefont{A.~N.} \bibnamefont{Chantis}},
  \bibinfo{author}{\bibfnamefont{M.}~\bibnamefont{van Schilfgaarde}},
  \bibnamefont{and} \bibinfo{author}{\bibfnamefont{T.}~\bibnamefont{Kotani}},
  \bibinfo{journal}{Phys. Rev. B} \textbf{\bibinfo{volume}{76}},
  \bibinfo{pages}{165126} (\bibinfo{year}{2007}).

\bibitem[{\citenamefont{Methfessel et~al.}(2000)\citenamefont{Methfessel, van
  Schilfgaarde, and Casali}}]{markbook}
\bibinfo{author}{\bibfnamefont{M.}~\bibnamefont{Methfessel}},
  \bibinfo{author}{\bibfnamefont{M.}~\bibnamefont{van Schilfgaarde}},
  \bibnamefont{and} \bibinfo{author}{\bibfnamefont{R.~A.}
  \bibnamefont{Casali}}, \emph{\bibinfo{title}{Lecture Notes in Physics}}, vol.
  \bibinfo{volume}{535} (\bibinfo{publisher}{Springer-Verlag, Berlin, 2000},
  \bibinfo{year}{2000}).

\bibitem[{\citenamefont{Kotani and van Schilfgaarde}(2002)}]{Kotani02}
\bibinfo{author}{\bibfnamefont{T.}~\bibnamefont{Kotani}} \bibnamefont{and}
  \bibinfo{author}{\bibfnamefont{M.}~\bibnamefont{van Schilfgaarde}},
  \bibinfo{journal}{Solid State Commun.} \textbf{\bibinfo{volume}{121}},
  \bibinfo{pages}{461} (\bibinfo{year}{2002}).

\bibitem[{\citenamefont{Blaha et~al.}(2001)\citenamefont{Blaha, Schwarz,
  Madsen, Kvasnicka, and Luitz}}]{wien2k-01}
\bibinfo{author}{\bibfnamefont{P.}~\bibnamefont{Blaha}},
  \bibinfo{author}{\bibfnamefont{K.}~\bibnamefont{Schwarz}},
  \bibinfo{author}{\bibfnamefont{G.~K.~H.} \bibnamefont{Madsen}},
  \bibinfo{author}{\bibfnamefont{D.}~\bibnamefont{Kvasnicka}},
  \bibnamefont{and} \bibinfo{author}{\bibfnamefont{J.}~\bibnamefont{Luitz}},
  \emph{\bibinfo{title}{WIEN2k, An Augmented Plane Wave + Local Orbitals
  Program for Calculating Crystal Properties}} (\bibinfo{publisher}{Karlheinz
  Schwarz, Techn. Universit¬at Wien, Austria, ISBN 3-9501031-1-2},
  \bibinfo{year}{2001}).

\bibitem[{\citenamefont{Luttinger}(1960)}]{Luttinger}
\bibinfo{author}{\bibfnamefont{J.~M.} \bibnamefont{Luttinger}},
  \bibinfo{journal}{Phys. Rev.} \textbf{\bibinfo{volume}{119}},
  \bibinfo{pages}{1153} (\bibinfo{year}{1960}).

\bibitem[{\citenamefont{Baer and Lang}(1980)}]{Baer}
\bibinfo{author}{\bibfnamefont{Y.}~\bibnamefont{Baer}} \bibnamefont{and}
  \bibinfo{author}{\bibfnamefont{K.}~\bibnamefont{Lang}},
  \bibinfo{journal}{Phys. Rev. B} \textbf{\bibinfo{volume}{21}},
  \bibinfo{pages}{2060} (\bibinfo{year}{1980}).

\end{thebibliography}
\end{document}